\documentclass[aps,prd,nofootinbib,twocolumn,superscriptaddress,showpacs,preprintnumbers]{revtex4-1}
\usepackage{epsfig,dcolumn}
\usepackage{graphicx}
\usepackage{color}

\usepackage{amsmath}
\usepackage{amsfonts}
\usepackage{amssymb}
\usepackage{graphicx}
\usepackage{bm} % scaled bold math symbols

\newcommand{\be}{\begin{equation}}
\newcommand{\ee}{\end{equation}}

\begin{document}

\title{Analytical improvements to the Breit-Wigner isobar models} 

%\author{V. Mathieu$^{1,2}$, G. Fox$^{3}$  and A. P. Szczepaniak$^{1,2,4}$} 
%\affiliation{
% $^1$ Department of Physics, Indiana University, Bloomington, IN 47405, USA \\
% $^2$  Center for Exploration of Energy and Matter, Indiana University, Bloomington, IN 47403, USA\\
% $^3$ School of Informatics and Computing, Indiana University, Bloomington, IN 47405, USA \\
 %$^4$  Jefferson Laboratory, 12000 Jefferson Avenue, Newport News, VA 23606, USA } 

\author{Adam~P.~Szczepaniak}
\affiliation{Physics Department, Indiana University, Bloomington, IN 47405, USA}
\affiliation{Center for Exploration of Energy and Matter, Indiana University, Bloomington, IN 47403}
\affiliation{Thomas Jefferson National Accelerator Facility, %12000 Jefferson Avenue,  
Newport News, VA 23606, USA}
\preprint{JLAB-THY-15-2149}

\begin{abstract}
We discuss the derivation and properties of the general representation of partial wave amplitudes in the context of improving the models currently used in analysis of three particle Dalitz distributions. 
\end{abstract}

%\pacs{11.55.Fv, 11.55.Hx, 12.40.Nn, 13.75.Gx}

\maketitle

\section{introduction}
In this note, after a brief introduction to aspects of  $S$-matrix theory 
 relevant in analysis of three particle Dalitz plots, I focus on properties of Breit-Wigner (BW) amplitudes and the isobar model in general.  I discuss the LHCb analysis model in the context of 
 a general  isobar-type approximation and  show, for example, 
   which features of the  BW amplitude, {\it e.g.} barrier factors,  Blatt-Weisskopf factors, {\it etc.} are universal    and which are not,  {\it i.e.}  are process dependent.  The possibility of extending the BW description in a way that is consistent with analyticity, unitarity, and even crossing 
    would allow to access systematic uncertainties in data analysis. I concentrate on spinless particles. 
     Spin introduces kinematical complexities but  
    does not affect how unitarity, analyticity, and crossing are implemented, at least for a finite set of partial waves. 

 \section{Kinematical vs Dynamical Singularities} 
 
 We are interested in amplitudes describing a decay of a  quasi-stable particle $D$ with mass $M$ to 
  three  distinguishable particles  $A,B,C$  
  
\begin{equation} 
D  \to A + B + C   \label{D} 
\end{equation} 
 The decay amplitude depends on particle helicities,  $\lambda_i$, 
 $i=A,B,C,D$,   and  three Mandelstam invariants, which we define as $s = (p_A + p_B)^2$, $t = (p_B + p_C)^2$ and $u=(p_A + p_C)^2$. The invariants  are kinematically constrained by $s + t + u = \sum_i m_i^2$.  Analytical $S$-matrix theory states that, besides the decay channel, the same amplitude describes each of the three  two-to-two scattering processes, {\it i.e} the $s$-channel reaction $D + \bar C \to A + B$, (bar denotes an antiparticle) as well as the $t$ and $u$ channel scattering.  What this means in practice is the following. For each combination of helicities  there is an analytical function $A_{\lambda_i}(s,t,u)$ of the three complex Mandelstam variables and complex $M^2$, such that the three physical 
    scattering amplitudes and the decay channel amplitude correspond to the limit of $A_{\lambda_i}$ 
     when  $s,t,u,$ and $M$ approach the real axis in the physical domain of the corresponding  reaction. 
 This is the essence of crossing symmetry. 
 In general crossing mixes helicity amplitudes and leads  
  to complicated relations for helicity amplitudes. Furthermore,  
     helicity amplitudes have kinematical singularities in the Mandelstam variables. 
% I don't understand this sentence: These are enforced by absence of such singularities in a covariant 
%representation.
 Despite such complexities it is possible to come up with parameterizations of helicity amplitudes that take into account both  kinematical and dynamical constraints~\cite{MartinSpearman}. 
   On the other hand it is also useful to consider 
    the  covariant form  {\it i.e.} a representation of helicity amplitudes in terms of  Lorentz-Dirac factors 
 that describe wave functions of the free particles participating in the reaction. The advantage of the covariant representation is that the scalar functions multiplying all independent covariants are simply related by crossing and are free from kinematical singularities. At the end of the day one still needs the 
     helicity amplitudes for partial wave analysis. In ~\cite{Danilkin:2014cra} the reaction $\omega \to \pi^+\pi^-\pi^0$ was  studied and this example provides a good illustration of the issues discussed above.

The main postulate of relativistic reaction theory is that reaction amplitudes are analytical functions of kinematical variables. It follows from Cauchy's theorem  that an analytical function is fully determined by its domain of analyticity, {\it i.e.} location of singularities. Thus knowing amplitude singularities allows 
to determine the amplitude elsewhere, including the various physical regions. 
     In S-matrix theory it is assumed that all singularities 
   can be traced to unitarity. In absence of an explicit solution to the scattering problem in QCD, 
   analyticity and unitarity  provide the least model dependent description  of hadron scattering. 
    
  Unitarity operates in any of the Mandelstam variables. 
 In the  $s$-channel the physical domain is located on the positive real axis in the $s$-plane above the elastic threshold.    Unitarity makes amplitudes singular at each open channel threshold, the singularity being of the square-root type. The same happens in variables $u$ and $t$.  The amplitude  also has singularities in the variable $M$ since it represents an unstable channel. In studying a particular decay process, {\it e.g.} $J/\psi \to 3\pi$,  $M$ is fixed in a very narrow range (within the width of the $J/\psi$),  and dependence on $M$ is effectively fixed and its singularity structure irrelevant.

In general,  it is not known how to write an amplitude that has correct unitarity 
 constraints in two or more overlapping channels. The reason being that it is simple to implement 
  unitarity on partial waves where it is an algebraic constraint. A single partial wave in one channel corresponds to an infinite number of partial  waves in another one thus imposing unitarity on say both $s$ and $t$-channel partial waves simultaneously requires an infinite number of partial waves. Regge theory  extends the 
  concept of analytical continuation to the angular momentum variable of partial waves. Singularities in the angular  momentum plane  {\it e.g.} Regge poles, determine behavior of infinite sums of partial waves, thus Regge theory is used to implement  cross channel unitarity.

 In general amplitude singularities are known only in a limited domain but as long as they dominate in 
  a kinematical region of interest one may be able to construct a realistic amplitude model.  Amplitude models fall into two main categories. One is that of dual models, {\it e.g.} the Veneziano model and  the other is the isobar  model category,  {\it e.g.} the Khuri-Treiman model. Dual models attempt to incorporate $S$-matrix constraints directly on the full amplitude that depends on the Mandelstam variables. 
  Since isobar is synonymous with a partial wave, isobar models are models based on a (truncated) partial wave expansion. 

   In the following I will discus the isobar model in some detail since this is almost exclusively the model used at present in analyses of three particle Dalitz distributions.

 \section{The Breit-Wigner amplitude} 
 
 In the  Breit-Wigner formula, a partial wave, $f_l(s)$ is approximated by a pole in $s$ 
  located in the complex energy plane at $s_P = Re s_P - i Im s_P$, 
\begin{equation} 
f^{p}_l(s) \propto \frac{1}{s_P - s}  \label{p} 
\end{equation} 
 The real and imaginary parts are related to the mass,  $M = \sqrt{Re\, s_P}$  and the width, $\Gamma = Im\, s_P/M$  of a resonance. For comparison with experiment a reaction amplitude 
  is evaluated at physical, real values of  kinematical variables {\it e.g} when $s$ approaches
    the real axis from above. The contribution from a  BW  pole to a partial cross section, $\sigma_l$ is  proportional to 

\begin{equation} 
\sigma_l \propto \lim_{\epsilon \to 0} |f^p_l(s + i\epsilon) |^2 \propto \frac{1}{(s - M^2)^2 + (M \Gamma)^2}.  
\end{equation} 

Since the resonance pole is located in the complex s-plane, on the real s-axis,  where experimental data is taken, it produces a smooth variation in the cross section. The closer the pole is to  the real axis, {\it i.e.} the smaller the resonances width, the more rapid is the variation in cross section or event distribution. 
It is worth keeping in mind that once energy  is considered as a complex variable any variation of the reaction amplitude in the physical region can be traced to existence of singularities in complex energy plane 
  {\it e.g.} a pole as in the BW formula.

   \section{ Unitarity and the Breit-Wigner amplitude} 
   
 The BW formula of Eq.~(\ref{p}) is an analytical function of $s$ except for a single pole at $s=s_P$. How does unitarity constrain the BW pole? Resonance decay is possible because of open channels and it is unitarity that controls distribution of probability across decays.  It thus follows that unitarity must 
  constrain resonance decay widths and thus the imaginary part of the BW pole. But Eq.~(\ref{p})  is only an approximation to the ``true", unitary amplitude valid for $s$ near the position of the complex pole. 
 Since  unitarity operates in the physical domain, {\it i.e.} 
       on the real axis, the constraint of unitarity on the ``true" amplitude is lost in the pole approximation, {\it i.e.} at a finite distance from the real axis. In this case implementing unitarity is related to using 
       energy dependent widths. 

      Suppose the lowest mass open channel is a state of  particles $A$ and $B$ with threshold at $s=s_{th} = (m_A + m_B)^2$. In the mass range between $s_{th}$  and the first threshold, 
    unitarity constrains the ``true" amplitude to satisfy,  
   
   \begin{equation} 
   Im \hat f_l(s) = \hat t^*_l(s) \rho_l(s) \hat f_l(s)  \label{f} 
  \end{equation} 
  Here $\hat f_l(s)  =  f_l(s)/(a q)^{l}$  and $\hat t_l(s) = t_l(s)/(a q)^{2l}$ are the $s$-channel 
   reduced partial waves 
    representing production  of $AB$ in $D + \bar C \to A + B$ and elastic $A + B \to A + B$ scattering, respectively. Near threshold $q\to0$, with  $q$  being the relative momentum between $A$ and $B$ in the 
     $s$-channel center of mass frame,  partial waves vanish as $(aq)^l$,  where $a$ is given 
      by the position of the lowest mass singularity in the crossed channels, {\it i.e.} the range of interaction.  $\rho_l(s)$ is a known kinematical function describing the two-body phase space. It has 
      a  square-root  branch point at $s = s_{tr}$. As $s$ increases past the first inelastic threshold, $Im \hat f_l(s)$ receives a  ``kick" from another square-root type singularity form channel openings  
       and the {\it r.h.s} of Eq.~(\ref{f}) needs to be modified.  Eventually three and more particle channels open. 
        In practice unitarity is a useful constraint in a limited energy range that covers a small number of open channels  {\it e.g.} close to the elastic threshold. 
              Replacing, in Eq.~(\ref{f}),  $\hat f_l$  by $\hat t$ one obtains the unitarity relation for the elastic 
         $A + B \to A + B$ partial waves in the elastic region, 
   \begin{equation} 
   Im \hat t_l(s) = |\hat t(s)|^2 \rho_l(s) \label{t} 
  \end{equation} 
 Above the inelastic threshold the {\it r.h.s} of Eq.~(\ref{t}) should be modified as discussed above. 
  In case there is a finite number, $N$  of  relevant inelastic channels the unitarity condition can be expressed in a matrix form 
\begin{equation} 
Im \hat t_{l,ij}(s) = \sum_k  \hat t_{l,ik}^*(s) \rho_{l,k}(s) \hat t_{kj}(s) 
\end{equation} 
 where $\hat t_{l,ij}(s) = t_l(s)/(aq)^{l_i}(aq)^{l_j}$  and 
 $\rho_{l,k}(s)$ is the appropriate, reduced phase space in the channel $k$. 
 The unitarity relation for $\hat f_{l,i}(s)$ takes on a similar form, 
\begin{equation} 
Im \hat f_{l,i}(s) = \sum_k  \hat t_{l,ik}^*(s) \rho_{l,k}(s) \hat f_{k}(s) 
\end{equation} 

For a given $\hat t_l(s)$,  the analytical amplitude $\hat f_l(s)$ that satisfies Eq.~(\ref{f}) can be written as  

    \begin{equation} 
  \hat f_l(s) = \hat t_l(s) G_l(s).  \label{sf} 
  \end{equation} 
and sometimes the so-called Muskhelishvili-Omnes function is used instead of $\hat t_l(s)$ on the {\it r.h.s} of Eq.~(\ref{sf})   \cite{Pham:1976yi}.  The function $G_l(s)$ is an analytical function of $s$ with cuts 
 except on the real axis in the elastic region. The latter are accounted for by the elastic amplitude $t_l(s)$. 
  The singularities  of $G_l(s)$ correspond to the (often unknown) contributions from the unitarity-demanded left hand cuts cuts that exist in the crossed, $t$ and $u$ channels. 
 The inelastic contributions {\it i.e.}  right cuts in $\hat f_l$ are related to the inelastic contributions to the amplitude    $\hat t_l(s)$ which is easy to show if the matrix representation is used, 
   \begin{equation} 
   \hat f_{l,i} = \sum_k \hat t_{l,ik} G_{l,k}(s) \label{sf2}
   \end{equation} 
 and with the functions $G_{l,k}(s)$ bearing only left hand cuts.    
     Now we go  back to the pole formula. From Eq.~(\ref{f})  it follows that poles of the production amplitude  
     $\hat f_l$ are also poles of $\hat t_l$. This is because no resonance poles appear on sheets connected 
   to the left hand cuts~\cite{gribov}.  Since $\hat t_l(s) $  satisfies  Eq.~(\ref{t}) it can be shown that the most general parametrization has the form, 
  \begin{equation} 
  t_l(s) = \frac{1 }{ C_l(s)  - I_l(s)  } \label{st} 
  \end{equation}
  which in the inelastic case generalizes to the matrix form 
  \begin{equation} 
  t_{l,ij}(s) = [ C_{l}(s) - I_l(s) ]^{-1}_{ij} 
  \end{equation}
 with $C_l$ and $I_l$ becoming   $N\times N$ matrices in the channel space.   
     The function $C_l(s)$ ($C_{l,ij}(s)$)   has  similar properties to the function $G_l(s)$  ($G_{l,i}(s)$)  in Eq.~(\ref{sf}),     {\it i.e.} they are real for real $s$ in the elastic region and have only left hand cuts. 
   The function $I_l(s)$ is a known analytical function {\it i.e.} the Chew-Mandelstam function, 
  %given by 
%\begin{equation}  
 %I(s) =   \frac{s}{\pi} \int ds' \frac{\rho(s')}{s' (s' - s)} 
 %\end{equation} 
with it's imaginary part for real $s$ given by $Im I_l(s) = \rho_l(s)$. 
The reason why the analytical solution of Eq.~(\ref{t}) is more complicated than that of Eq.~(\ref{f}), is that the former is a non-linear relation for the amplitude. It is easy to check that this equation becomes a linear condition for the inverse of $\hat t_l$  
 and this is the reason why dependence on phase space appears in the denominator in Eq.~(\ref{st}). 
 It is straightforward  to check that Eq.~(\ref{st}) satisfies Eq.~(\ref{f}), or in the inelastic case,  its matrix generalization.  Presence of  ``denominators" in amplitudes are a direct consequence of unitarity and so are the resonance  poles, which correspond to zeros of the denominators.

  One  immediately recognizes that the $K$-matrix, or the $K$ function in the elastic case 
   corresponds to 
 \begin{equation} 
 K^{-1}_l(s) = C_l(s) - Re I_l(s). 
 \end{equation} 
 Since the real  part of a function is not an analytical function an analytical approximation to the $K$ matrix 
  violates analyticity of the amplitude and may lead to spurious ``kinks" from square-root unitarity branch points in the physical region. 
 It  is much better to use Eq.~(\ref{st}), {\it aka} the Chew-Mandelstam representation, with the analytical ($K$-matrix type) parametrization reserved for the analytical  function $C_l(s)$. 
       
  Combining Eq.~(\ref{st}) with Eq.~(\ref{sf}) one obtains 
  \begin{equation} 
  \hat f_l(s) = \frac{G_l(s)}{C_l(s) - I_l(s)}  \label{solf1} 
  \end{equation} 
  or in the matrix form for the inelastic case 
  \begin{equation} 
  \hat f_{l,i}(s) = \sum_k [C_l(s) - I_l(s)]^{-1}_{i,k} G_{l,k}  \label{solf2} 
  \end{equation}

  Now we can finally see how  the BW  pole formula of Eq.~(\ref{p})  emerges. Suppose in Eq.~(\ref{solf1}) the denominator vanishes  at some complex $s= s_p$. Near the pole  of  $\hat f_l$ 

\begin{equation} 
 \hat  f_l(s) \sim \frac{\beta_l}{s - s_p}  
  \end{equation} 
  where 
  \begin{equation} 
  \beta_l = G_l(s_p)/(C'_l(s_p) - I'_l(s_p))
  \end{equation} 
  In the inelastic case the role of the denominator in the {\it r.h.s} of  Eq.~(\ref{solf1}) is  played by the  determinant of the $N \times N$ matrix $[C_l(s) - I_l(s)]$. 
  Even though the residue of the pole, $\beta_l$  is in general is a complex number, it can be shown that only its magnitude is to be related with  a coupling of a resonance to a decay channel~\cite{gribov}. The residue $\beta_l$ should be distinguished from the numerator $G_l(s)$. The latter is energy dependent and represents production amplitude of the final state  $A + B$ given the initial state $D + \bar C$. The former is a number representing the  product of couplings of the resonance to the  initial and final states. 
     
 Finally we  ``derive" the more familiar  BW formula, with energy dependent widths. In  the ``LHCb" notation 
 \begin{equation} 
 f^{LHCb}_{l_s}(s) = F_{l_s}(q) R_{l_s}(s) F_{l_t}(p)  
 \end{equation} 
 so that the reduced amplitude is given by 
 \begin{equation} 
 \hat f^{LHCb}_{l_s}(s) = \frac{ F_{l_s}(q)}{(aq)^{l_s}} R_{l_s}(s) F_{l_t}(p)
 \end{equation} 
 where $q = q(s)$ is the decay channel relative momentum between $A$ and $B$ 
  and $p = p(s)$ is the decay channel relative momentum between the $(AB)$ pair and the spectator particle $C$. For comparison with the analysis given above all is needed is to replace the decay channel expressions for $q$ and $p$ by the $s$-channel ones. 
 The function $F_l(x)$ is a product of an angular momentum barrier factor  $x^{l}$ and a Blatt-Weisskopf  factor   
 \begin{equation} 
 F_l(x) = (ax)^l F'_l(x)
 \end{equation} 
where, for example, 
 \begin{equation} 
 F'_2(q) =  \sqrt{\frac{13}{((aq)-3)^2 + 9(aq)}}  
 \end{equation} 
 The propagator $R_{l_s}(s)$ is given by 
 \begin{eqnarray} 
& &  R_{l_s}(s) = \frac{1}{ m_r^2 - s - i \rho_{l_s}(s) X(p) } \nonumber \\
 & &  = 
  \frac{1}{  X(p) (  C^{LHCb}(s) -  i \rho_{l_s}(s) ) }, \, C^{LHCb}(s)   \equiv \frac{m_r^2 - s}{X(p)}  \nonumber \\
 \end{eqnarray} 
 so that one can rewrite the LHCb amplitude model as 
 \begin{equation} 
 \hat f^{LHCb}_{l_s}(s) = \frac{G^{LHCb}_{l_s}(s)}{ C^{LHcB}(s) - i\rho_{l_s}(s) }  \label{BWlhcb} 
 \end{equation} 
 where 
 \begin{equation} 
 G^{LHCb}_{l_s}(s)   =  \frac{F'_{l_s}(q) F_{l_t}(p)}{X(s)} 
 \end{equation} 
 Eq.~(\ref{BWlhcb}) is a specific case of Eq.~(\ref{solf1}). It is in fact the $K$ matrix (function) approximation 
 since only the imaginary part, $\rho_{l_s}(s)$ of the dispersive integral $I_l(s)$ is used.
   The functions $C_l$ and $G_l$ containing, through left hand cuts,  physics of elastic production of $AB$ in  $A + B \to A + B$ and  in decay $D  \to A + B + C$, respectively, 
    have been replaced by a specific product of Blatt-Wisskopf factors. The latter originate 
     from a potential model in non-relativistic scattering and at best can be considered as a crude  approximation. In precision data analysis they should be replaced by a more flexible 
       parametrization.  In LHCb analysis contribution from several poles are included by adding 
        BW amplitudes. Eq.~(\ref{solf1}) shows how all such poles need to appear as zeros of the common denominator. Finally the matrix representation of Eq.~(\ref{solf2})  is  the correct formula for dealing with multiple channels.

   \section{ Combining $s$, $t$, and $u$, channel isobars and corrections to the isobar model} 
 
In the previous section we discussed  how the energy dependent  BW amplitude is related to the general 
 expression for the partial wave. Here we discuss how the partial wave amplitudes build the full amplitude in the decay channel. 
 
In the scattering domain of the $s$-channel  partial wave expansion of the full amplitude 
 converges and a finite number of partial waves waves may give a good approximation to the whole sum.  Energy dependence of individual 
  partial wave can be represented using expressions like the one in Eq.~(\ref{sf}) or  Eq.~(\ref{sf2}). 
  In a decay channel partial wave series also  converges, but one cannot simply replace the one sum but the other. This is because  the decay channel, partial waves have extra ''complexity'' compared to the scattering channel, due to $t$ and $u$ channel singularities, {\it i.e.} resonances begin in the physical region. 
     Ignoring spins of external particles,    the $s$-channel partial wave series is given by 
  \begin{equation} 
  A(s,t,u) = \frac{1}{4\pi} \sum_{l=0}^\infty (2l+1) f_l(s) P_l(z_s) \label{spw} 
  \end{equation} 
On the {\it r.h.s} the dependence on $t$ and $u$ is algebraic, through $z_s$ and the rotational functions. Thus in the physical domain of the decay the $s$-channel series diverges because the {\it l.h.s} has singularities in $t$ and $u$. Furthermore, it follows that any truncated, finite set of $s$-channel partial waves cannot reproduce $t$ or $u$-channel singularities, {\it e.g.} resonance which appear inside the Dalitz plot. These issues are resolved in the isobar model by replacing the infinite number of $s$-channel partial waves by a  finite set and  adding  a (finite)  set of $t$ and $u$ channel partial waves. Thus the amplitude has a mixed form that includes partial waves (isobars) in the three channels simultaneously, 
 
 \begin{equation}
 A(s,t,u) = \frac{1}{4\pi} \sum_{l=0}^{L_{max}} (2l+1) a^{(s)}_l(s) P_l(z_s)  + (s \to t) + (s \to u) \label{im} 
 \end{equation} 
 We refer to $a^{(s)}_l(s)$ as the isobaric amplitude in the $s$-channel and analogously, 
   $a^{(t)}_l(t)$ and $a^{(u)}_l(u)$ are the isobaric amplitudes in the $t$ and $u$-channels, respectively. 
 In a typical Dalitz plot analysis, the isobaric amplitudes are parametrized using the energy dependent BW amplitudes discussed in the previous sections.  
 
 Projecting the {\it r.h.s} of Eq.~(\ref{im}) into the $s$-channel gives the $s$-channel partial waves, which 
   we denoted by $f_l(s)$,  
 \begin{eqnarray} 
& &  f_l(s)  = a^{(s)}_l(s) \nonumber \\
& &  +\frac{1}{2} \int_{-1}^1  dz_s P_l(z_s) \sum_{l'=0}^{L_{max}} (2l' + 1) a^{(t)}_{l'}(t) P_{l'}(z_t) + (t \to u) \nonumber \\  & & \equiv a^{(s)}_l(s) + b^{(s)}_l(s). 
 \end{eqnarray} 
 Under the integral, $t$ and $u$ are to be considered as function of $s$ and $z_s$, the cosine of the $s$-channel scattering angle.  Since the integral contributes to partial waves with arbitrary $l$, Eq.~(\ref{im}) defines a model   for an infinite number of partial waves $f_l(s)$ and gives the result of the  analytical continuation of the   series in Eq.~(\ref{spw}).  
Application of unitarity in the $s$-channel leads to a relation between the isobaric  amplitudes 
\begin{equation} 
a^{(s)}_l(s) = t_l(s) \left[ \frac{1}{\pi} \int_{s_{tr}} ds' \frac{ \rho_l(s') b^{(s)}_l(s')}{s' - s} \right] \label{a} 
\end{equation} 
The amplitude  $b^{(s)}_l(s)$ is  the $s$-channel projection of the $t$ and $u$ channel exchanges. 
 As a function of $s$ it has the left hand cut but no right hand, unitary  cut. The 
dispersive integral in Eq.~(\ref{a})  has the $s$-channel unitary cut. 
 Thus $s$-channel unitarity demands that $s$-channel isobaric amplitude $a^{(s)}_l(s)$ has the right cut 
  coming not only form the elastic $A + B \to A + B$ amplitude, $t_l(s)$, but also from dispersion of $s$-channel projections of the the  $t$ and $u$-channel amplitudes. It is important to note that difference between the partial wave amplitudes $f_l(s)$ and the isobaric amplitudes $a^{(s)}_l(s)$. In case of the former the right hand cut discontinuity comes entirely  from the elastic scattering, {\it c.f.} Eq.~(\ref{sf}),~(\ref{sf2}). In the isobar model, the partial wave amplitudes are given by, 
\begin{eqnarray}  
& &  f_l(s) = t_l(s) \left[ \frac{ b^{(s)}_l(s) }{t_l(s)}+  \frac{1}{\pi} \int_{s_{tr}} ds' \frac{ \rho(s') b^{(s)}_l(s')}{s' - s} \right]  \nonumber \\
& & \equiv  t_l(s) G_l(s)  
\end{eqnarray} 
which are indeed of the form given by Eq.~(\ref{sf}), since it can be shown that the right hand cuts cancel between the two terms in the square bracket so that  $G_l(s)$ has only left hand cuts,

\begin{equation}  
  G_l(s) =  \frac{1}{\pi} \int_{s_{tr}} ds' \rho(s') \frac{  b^{(s)}_l(s') - b^{(s)}_l(s)}{s' - s} 
\end{equation}

 In the inelastic case it generalizes to the form given by Eq.~(\ref{sf2}). 
 The left hand cuts, as expected, originate from exchanges in the crossed channels determined by the amplitude $b^{(s)}_l(s)$. 
  
   One goes further and also  impose $t$ and $u$-channel unitarity, thereby correlating isobar expansions in the three channels. This is the analytical, unitary description of ``final state interaction"   that relies on the model independent features of the amplitudes only. 
 
  The implications for event distributions  in the Dalitz plot, given the left hand cut singularities of 
  $b^{(s)}_l(s)$ were studied in ~\cite{schmid} and in the next section we take a  look at the left hand cuts of the amplitude $G_l(s)$.

  \section{Watson's theorem} 
  Since the imaginary part of a complex function is itself a real function, it follows from Eq.~(\ref{f}) that in the elastic region, phase of $\hat f_l(s)$ equals that of the elastic scattering amplitude $\hat t_l(s)$. 
It is often forgotten to be mentioned, however, that in many cases the {\it r.h.s}  in 
 Eq.~(\ref{f}) does not saturate the imaginary part even in the elastic regime. When this happens, Watson's theorem is violated even in the elastic regime. And this is the case of a decay process. In this case there is nevertheless a relation between $\hat f_l$ and $\hat t_l$, or a generalized Watson relation. 
 
 In the decay kinematics, even if the energy of the $A B$ state is below inelastic threshold, the imaginary part of the $s$-channel partial amplitude, $\hat f_l(s)$  is not in given by Eq.~(\ref{f}). This happens because  in the decay kinematics cross channel singularities (from thresholds/unitarity) {\it e.g.} isobar exchanges in $t$ or $u$ channel are located in the physical region of the $s$-channel and contribute to the imaginary part of the  $s$-channel partial wave. Another way of saying this, is that in the decay kinematics singularities of $\hat f_l(s)$ which otherwise are to the left of the  elastic unitarity cut move to the right of the elastic unitary branch point.  
 
In the previous section we argued that  Eq.~(\ref{sf}) follows from the assumption that $\hat f_l(s)$ is an analytical function.  The relation between analyticity and casualty applies to full amplitudes, while partial waves are related to full amplitudes in a complicated way. Analytical partial waves are  obtained by a continuation of the unitary relation to the complex energy plane. It can be shown  (see {\it e.g.} \cite{gribov}) 
  that when left and right hand cuts are separated $Im \hat f_(s)   = \Delta \hat f(s_+) \equiv  (\hat f_l(s_+) - \hat f_l(s_-) )/2i $, where $s_\pm = s \pm  i \epsilon $. That is,  the imaginary part of the amplitude as measured in the experiment, which by itself is a real function, is  equal to the discontinuity across the real axis of the unique extension of $\hat f_l(s)$ to the complex energy plane. It turns out,  however that in decay kinematics, the proper extension of $\hat f_l(s)$ to the complex energy plane is such that in 
  Eq.~(\ref{f}) the {\it l.h.s} should be replaced by 
 $\Delta f(s_+)$ and the {\it r.h.s} should be replaced by the product 
 $t_l(s_-) \rho_l(s) f_(s_+)$.  Thus, elastic unitarity still determines the discontinuity across the right hand cut but it is no longer a real function. This is the generalized Watson or discontinuity relation.

 As a result the representation given by Eqs.~(\ref{sf}),(\ref{sf2}) is still valid 
with  the exception that the numerators, $G_l(s)$ become complex in the elastic region. They do not have the unitarity cut though. Just like in the ``standard" Watson's theorem the right hand cut discontinuity comes entirely from the elastic amplitude $t_l(s)$.  The complexity of $G_l(s)$ is still of the left hand cut nature, except that left hand cuts have moved onto the right hand side under the right hand cut and into the second sheet. 
\cite{Szczepaniak:2015eza}.

\section*{Acknowledgments}
I would like to thank the organizers and participants of the LHCb workshop on multi-body decays of B and D mesons for stimulating discussions. 
 This  work is  supported by the U.S. Department of Energy, Office of Science, Office of Nuclear Physics under contract DE-AC05-06OR23177. It is  also supported in part by the U.S. Department of Energy under Grant No. DE-FG0287ER40365.

  \end{document}